\documentclass[preprintnumbers,amsmath,amssymb,floatfix,9pt,prd,twocolumn,
superscriptaddress,nofootinbib]{revtex4}
\usepackage{latexsym}
\usepackage{epsfig}
\usepackage{amssymb} 
\begin{document}

\title{A charged anisotropic well-behaved Adler-Finch-Skea solution Satisfying Karmarkar Condition}

\author{Piyali Bhar}
\email{piyalibhar90@gmail.com
 } \affiliation{Department of
Mathematics,Government General Degree College, Singur, Hooghly 712 409, West Bengal,
India}

\author{Ksh. Newton Singh}
\email{ntnphy@gmail.com}
\affiliation{Department of Physics, National Defence Academy, Khadakasla, Pune-411023, India.}

\author{Farook Rahaman}
\email{rahaman@associates.iucaa.in}
\affiliation{Department of Mathematics, Jadavpur University, Kolkata, West Bengal-700032, India.}

\author{Neeraj Pant}
\email{neeraj.pant@yahoo.com}
\affiliation{Department of Mathematics, National Defence Academy, Khadak- wasla, Pune-411023, India.}
\author{Sumita Banerjee }
\email{banerjee.sumita.jumath@gmail.com}
\affiliation{Department of Mathematatics,Budge-Budge Institute of Technology,Budge-Budge, West-Bengal, India.}

\begin{abstract}\noindent
In the present article, we discover a new well-behaved charged anisotropic solution of Einstein-Maxwell's field equations. We ansatz the metric potential $g_{00}$ of the form given by Maurya el al. (arXiv:1607.05582v1) with $n=2$. In their article it is mentioned that for $n=2$ the solution is not well-behaved for neutral configuration as the speed of sound is non-decreasing radially outward. However, the solution can represent a physically possible configuration with the inclusion of some net electric charged i.e. the solution can become a well-behaved solution with decreasing sound speed radially outward for a charged configuration. Due to the inclusion of electric charged the solution leads to a very stiff equation of state (EoS) with the velocity of sound at the center $v_{r0}^2=0.819, ~v_{t0}^2=0.923$ and the compactness parameter $u=0.823$ is closed to the Buchdahl limit 0.889. This stiff EoS support a compact star configuration of mass $5.418M_\odot$ and radius of $10.1 km$.\\

pacs: 02.60.Cb;  04.20.-q;  04.20.Jb;  04.40.Nr;  04.40.Dg
\end{abstract}

\maketitle
\section{Introduction}

Many studies on astrophysical massive compact objects assumed the matter distribution is generally isotropic. However, such simplified assumptions yields satisfactory results to some extend and not for all systems. Recent researches in theoretical physics on the compact stellar systems suggested that matter distribution at the interior of these compact objects is most probably to be anisotropic from certain density ranges \cite{rud,can}. In the light of these studies, a new physics emerges to study the properties of anisotropic matter distributions in general relativity. The anisotropy in pressure could be introduced by the existence of a solid core, type P superfluid, complex nuclear interactions, or inclusion of net electric charge. The energy-momentum tensor $T^{\mu \nu}$ of such anisotropic matter is equivalent to those by assuming a fluid composed of two perfect fluids, or a perfect fluid and a null fluid, or two null fluids \cite{let1,let2,let3,bay}. \par
Most of the method commonly adopted by many researchers to explore new analytic solutions of the field equations are by assuming  $g_{00}$ or $g_{11}$, radial pressure, anisotropy, electric field, density and equation of state so that the field equations are integrable \cite{dev1,dev2,esc,mah1,mah2,ntn1,ntn2,ntn3,ntn4,maur1,maur2}. The Buchdahl limit of ideal fluid distributions postulated that the compactness parameter $u=2M/R$ should be $\le 8/9$  so that it doesn't proceed a gravitational collapse so that can form a singularity or Black Hole. This upper bound in $u$ is generalized by Andreasson \cite{and1,and2} with the inclusion of charge, anisotropy and even cosmological constant. In the derivation of the new generalized upper bound in $u$, he assumed a simple inequality in between pressure and density as $p_r+2p_t\le \rho$.\par
Many investigations have also shown that the behavior of a collapsing star is strongly influence by its initial static configuration on account for various parameters like, pressure anisotropy, charge, EoS, shear, radiations etc. Two static configurations at the initial with same masses and radii for different pressure profiles when undergoing collapse leads to very different temperature evolution at their later stages \cite{nai}. Hence, to completely understand the physics evolving stars one needs to understand the initial static configurations with the inclusion these various factors.  In Newtonian  approximation, adiabatic collapse of a fluid distribution obeying polytropic EoS is possible only when the adiabatic index $\Gamma < 4/3$. However, this idea is turnover by Chan et al. \cite{chan} for anisotropic fluid objects where collapse is still possible even for $\Gamma \ge 4/3$ depending on the nature of anisotropy. Herrera and Santos \cite{her1} have also suggested that the stability of static stellar configurations can be enhanced by the nature of local anisotropy. Inclusion of charge is the source of electric field which can cause pressure anisotropy Usov \cite{uso} and it can counter balanced the gravitational attraction by the electric repulsion other than the pressure gradient. On using this concepts Ivanov \cite{iva} proposed a model for charged perfect fluid that inhibits the growth of spacetime curvature to avoid singularities. Bonnor \cite{bon} also pointed out that a dust distribution of arbitrarily large mass can be  bounded in very small radius and maintained equilibrium against the gravity by a repulsive force produced by a small amount of charge. Thus it is interesting to study the implications of Einstein-Maxwell's field equations in general relativistic.\par
Bhar et al. \cite{bhar1} also presented new class of exact interior solution of Einstein-Maxwell's field equations in $(2 + 1)$ dimensional spacetime by assuming Chaplygin gas EoS and Krori-Barua metric with charged BTZ spacetime as exterior. Using this solution they have discussed all the physical properties of charged anisotropic stellar configuration. Bhar and Rahaman \cite{bhar2} proposed a new model of dark energy star consisting of five zones viz solid core of constant energy density, thin shell between core and interior, an inhomogeneous interior region with anisotropic pressures, a thin shell, and the exterior vacuum region. Bhar \cite{bhar3} also used the Krori-Barua metric potentials in the presence of quintessence field to model stable strange star model.\par
In this article, we have adopted a new method in order of discover new exact solution of Einstein-Maxwell's field equations that satisfies Tolman-Oppenheimer-Volkoff (TOV) equation. We adopted the method used by Karmarker 1948 to solve the field equation where the obtained solutions are classified as Class One. In this method the Reimann curvature tensor $\mathcal{R}_{\mu \nu \alpha \beta}$ satisfy a particular equation that finally link the two metric component $g_{00}$ and $g_{11}$ in a single equation i.e. the two metric components are dependent on each other. Therefore, we only need to assume one of the metric potential and electric field intensity to integrate the field equations. The rest of the physical quantities like pressure, density, sound speed, anisotropy, etc. can be completely determine from $g_{00},~g_{tt}$ and $E^2$ only. Many other articles are also available in literature on embedding Class One solutions \cite{kn1,kn2,kn3,kn4,kn5,kn6,kn7,bhar4}.

\section{Basic field equations}
To describe the interior of a static and spherically symmetry object the line element can be taken in canonical co-ordinate as,
\begin{equation}
ds^{2}=-e^{\nu(r)}dt^{2}+e^{\lambda(r)}dr^{2}+r^{2}\left(d\theta^{2}+\sin^{2}\theta d\phi^{2} \right) \label{met}
\end{equation}
Where $\nu$ and $\lambda$ are functions of the radial coordinate `$r$' only.\\

Now if the space-time (\ref{met}) satisfies the Karmarkar condition \cite{kar}
\begin{equation}
\mathcal{R}_{1414}\mathcal{R}_{2323}=\mathcal{R}_{1212}\mathcal{R}_{3434}+ \mathcal{R}_{1224}\mathcal{R}_{1334} \label{kar}
\end{equation}
with $\mathcal{R}_{2323}\neq 0$ \cite{pandey}, it represents the space-time of emending class 1.\\

For the condition (\ref{kar}), the line element (\ref{met}) gives the following differential equation
\begin{equation}
\frac{\lambda'\nu'}{1-e^{\lambda}}=-2(\nu''+\nu'^{2})+\nu'^{2}+\lambda'\nu'. \label{diff}
\end{equation}
with $e^{\lambda}\neq 1.$ Solving equation (\ref{diff}) we get,
\begin{equation}
e^{\lambda}=1+F\nu'^{2}e^{\nu} \label{elam}
\end{equation}
where $F \neq 0$, an arbitrary constant.\\

We assume that the matter within the star is charged and anisotropic in nature. The corresponding the energy-momentum tensor is described by,
\begin{eqnarray}
T^\mu_\xi &=& \rho v^\mu v_\xi + p_r \chi_\xi \chi^\mu + p_t(v^\mu v_\xi -\chi_\xi \chi^\mu-g^\mu_\xi) \nonumber\\
&&+{1\over 4\pi}\left(-\mathcal{F}^{\mu \nu} \mathcal{F}_{\xi \nu}+{1\over 4}g^\mu_\xi \mathcal{F}_{\sigma \nu} \mathcal{F}^{\sigma \nu}\right)\label{ten}
\end{eqnarray}
here all the symbols have their usual meanings.\\

Now for the line element (\ref{met}) and the matter distribution (\ref{ten}) Einstein-Maxwell's Field equations (assuming $G=c=1$) take the form,
\begin{eqnarray}
\mathcal{R}^\mu_\xi-{1\over 2}\mathcal{R}~g^\mu_\xi &=& -8\pi T^\mu_\xi \label{ein}\\
{1\over \sqrt{-g}}{\partial \over \partial x^\beta}\left(\sqrt{-g}~\mathcal{F}^{\mu \beta}\right) &=& -4\pi \mathcal{J}^\mu \label{max1}\\
\mathcal{F}^{\mu \nu}_{;\beta}+\mathcal{F}^{\nu \beta}_{;\mu}+\mathcal{F}^{\beta \mu}_{;\nu} &=& 0 \label{max2}
\end{eqnarray}

The Maxwell's Stress Tensor $\mathcal{F}^{\mu \beta}$ is defined by
\begin{equation}
\mathcal{F}^{\mu \beta}=\partial^\beta A^\mu -\partial^\mu A^\beta
\end{equation}

Here $A^\mu=(\phi(r),0,0,0)$ is the magnetic four potential and $\mathcal{J}^\mu$ is four-current density defined as
\begin{equation}
\mathcal{J}^\mu={\sigma_0 \over \sqrt{g_{00}}}{dx^\mu \over dx^0}
\end{equation}
provided $\sigma_0$ is the proper charge density.\\

For a static fluid configuration, the non-zero components of the four-current density is $j^0$ and function of $r$ only because of spherical symmetry. From (\ref{max1}) we get
\begin{equation}
\mathcal{F}^{01}=-e^{-(\nu+\lambda)/ 2} ~~{q(r) \over r^2}
\end{equation}
where $q(r)$ is the charge enclosed within a sphere of radius $r$ and given by
\begin{equation}
q(r)=4\pi \int_0^r e^{\lambda/2}\sigma_0 \eta^2 d\eta
\end{equation}

The Einstein-Maxwell's field equations (\ref{ein})-(\ref{max2}) reduces to 4-system of non-linear differential equations given by,
\begin{eqnarray}
\frac{1-e^{-\lambda}}{r^{2}}+\frac{e^{-\lambda}\lambda'}{r} &=& 8\pi\rho+E^{2} \label{dens}\\
\frac{e^{-\lambda}-1}{r^{2}}+\frac{e^{-\lambda}\nu'}{r} &=& 8\pi p_{r}-E^{2}\\
e^{-\lambda}\left(\frac{\nu''}{2}+\frac{\nu'^{2}}{4}-\frac{\nu'\lambda'}{4}+\frac{\nu'-\lambda'}{2r} \right) &=& 8\pi p_t+E^{2} \label{prt}\\
{e^{-\lambda/2} \over 4\pi r^2} \Big(r^2E\Big)' &=& \sigma(r)	\label{char}
\end{eqnarray}
where $\sigma(r)$ is the charge density and $E=q(r)/r^2$ is the electric field intensity at the interior.\\

Now we have to solve the Einstein-Maxwell's field equations (\ref{dens})-(\ref{char}) with the help of equation (\ref{elam}). One can notice that we have five equations with 6 unknowns namely $\lambda,~\nu,~\rho,p_r$, $p_t$ and $E$. To solve the above set of equations let us ansatz the metric co-efficient $g_{tt}$ proposed by Adler \cite{adl} as,
\begin{equation}
e^{\nu}=B(1+Cr^{2})^{2} \label{enu}
\end{equation}
Where $B$ and $C$ are constants.\\

Let us assume the electric field as well in the form given below
\begin{equation}\label{e2}
E^{2}=\frac{KCr^{2}}{1+Cr^{2}}
\end{equation}

On using equation (\ref{elam}) and (\ref{enu}) we obtain,
\begin{equation}
e^{\lambda}=1 + 16 B C^2 F r^2
\end{equation}
 This metric form of $e^\lambda$  is similar to that of Finc-Skea solution \cite{finch}.

Now employing the values of $e^{\nu}$ and $e^{\lambda}$ to the Einstein-Maxwell's field equations (\ref{dens})-(\ref{char})and using the expression for $E^2$ given in eq.(\ref{e2}) we obtain the expression for matter density, radial \& transverse pressure and proper charge density as,
\begin{eqnarray}
8\pi\rho &=& \frac{16 B C^{2} F (3 + 16 B C^2 F r^2)}{(1 + 16 B C^2 F r^2)^2}-\frac{K C r^2}{1 + C r^2} \label{r1}\\
8\pi p_r &=& \frac{4C + K C r^2 - 16 B C^2 F \{1 + C r^2 (1 - K r^2)\}}{(1 + C r^2) (1 + 16 B C^2 F r^2)}\nonumber\\
\\
8\pi p_t &=& \frac{1}{(1 + C r^2) (1 + 16 B C^2 F r^2)^2}\Big[4C - K Cr^2 \nonumber\\
&&- 256 B^2 C^5 F^2 K r^6 -
16 B C^2 F \times\\
&&\{1 - C r^2 (1-2 K r^2)\}\Big] \nonumber\\\label{r2}\\
\sigma(r) &=& \frac{C K r (Cr^2+2)}{2 \pi  (C r^2+1)^2 \sqrt{16 B C^2 F r^2+1}} \label{r3}
\end{eqnarray}
and the anisotropic factor $\Delta$ is obtained as,

\begin{eqnarray}
8\pi \Delta &=& 8\pi(p_t-p_r)\nonumber\\
&&= \frac{2 C r^2} {(1 + C r^2) (1 + 16 B C^2 F r^2)^2}\times \big[-K \nonumber\\&&-
   16 B C^2 F \{1 + 2 K r^2 - 8 B C F
   (1 + C r^2- 2 K C r^4)\}\big] \nonumber\\\label{r4}
\end{eqnarray}

\section{Boundary Conditions and determination constants}

We match our interior space-time to the exterior Reissner- N$\ddot{o}$rdstrom line element given by
\begin{eqnarray}
ds^{2}&=&-\left(1-\frac{2m}{r}+\frac{q^2}{r^{2}}\right)dt^{2}+\left(1-\frac{2m}{r}+\frac{q^2}{r^{2}}\right)^{-1}dr^{2}\nonumber\\
&&+r^{2}(d\theta^{2}+\sin^{2}\theta d\phi^{2})
\end{eqnarray}
with the radial coordinate $r>m+\sqrt{m^{2}-q^2}$ \\

Using the continuity of the metric coefficient $e^{\nu}$ and $e^{\lambda}$ across the boundary we get the following equations
\begin{eqnarray}
1-\frac{2M}{r_b}+\frac{q^2(r_b)}{r_b^{2}} &=& B(1+C r_b^{2})^{2} \label{b1}\\
\left(1-\frac{2M}{r_b}+\frac{q^2(r_b)}{r_b^{2}}\right)^{-1} &=& 1 + 16 B C^2 F r_b^2   \label{b2}\\
p_r(r=r_b) &=& 0 \label{b3}
\end{eqnarray}

On using the boundary conditions (\ref{b1})-(\ref{b3}) we get
\begin{eqnarray}
F &=& \frac{K r_b^2+4}{16 C B (1-C K r_b^4+C r_b^2)} \\
B &=& \frac{C K r_b^5-2 C M r_b^2+C r_b^3+r_b-2 M}{r_b (C r_b^2+1)^2}\\
C &=& \frac{1}{2 (3 K r_b^7-5 M r_b^4+2 r_b^5)}\times \nonumber\\&&\Big[-r_b^2 \sqrt{K^2 r_b^6+4 K r_b^4-16 M r_b+4 r_b^2+16 M^2} \nonumber\\&&-K r_b^5+6 M r_b^2-2 r_b^3\Big]
\end{eqnarray}
We have chosen the $M,~r_b$ and $K$ as free parameters.

\begin{figure*}[thbp]
\begin{center}
\begin{tabular}{rl}
\includegraphics[width=6.8cm]{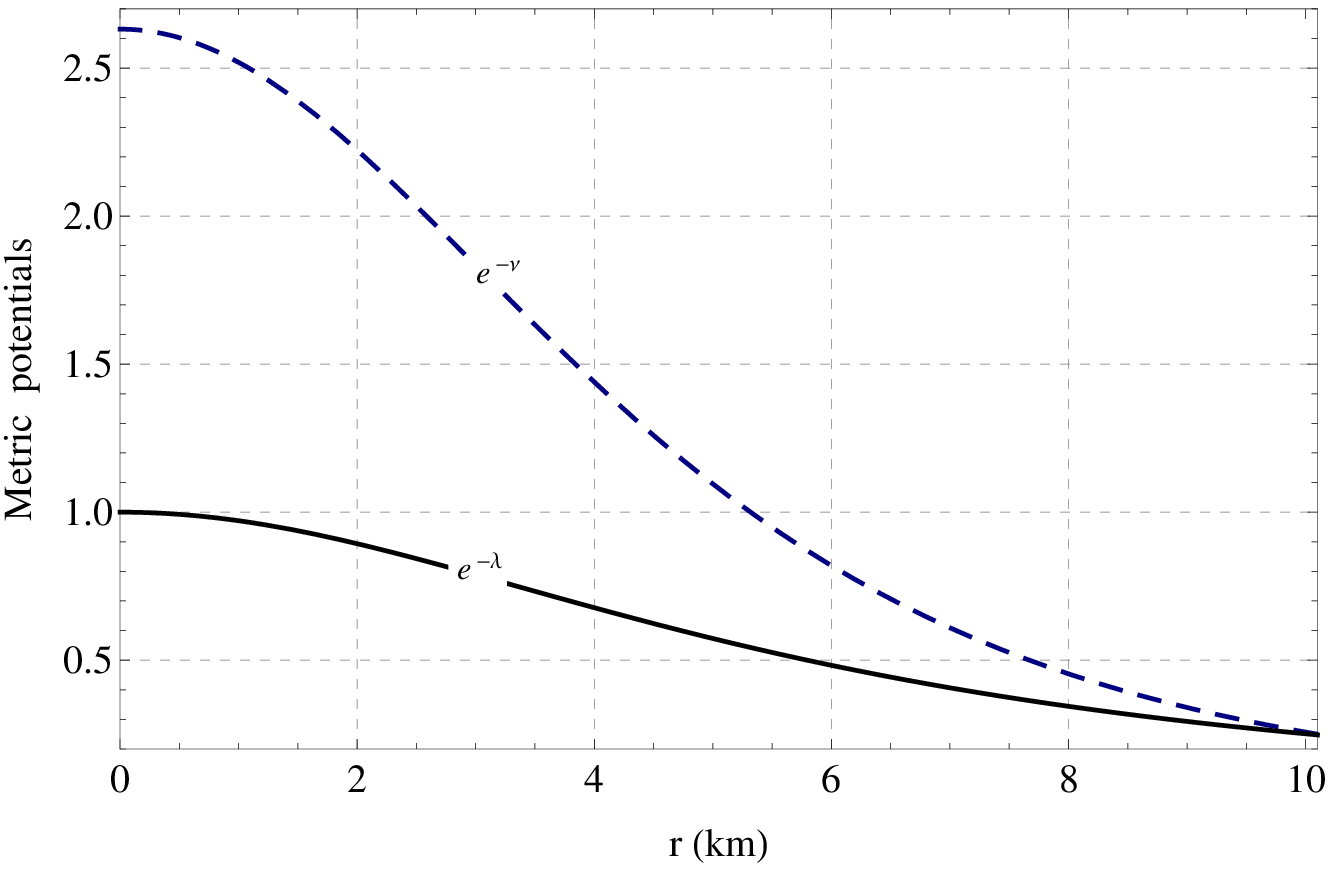}&
\includegraphics[width=6.8cm]{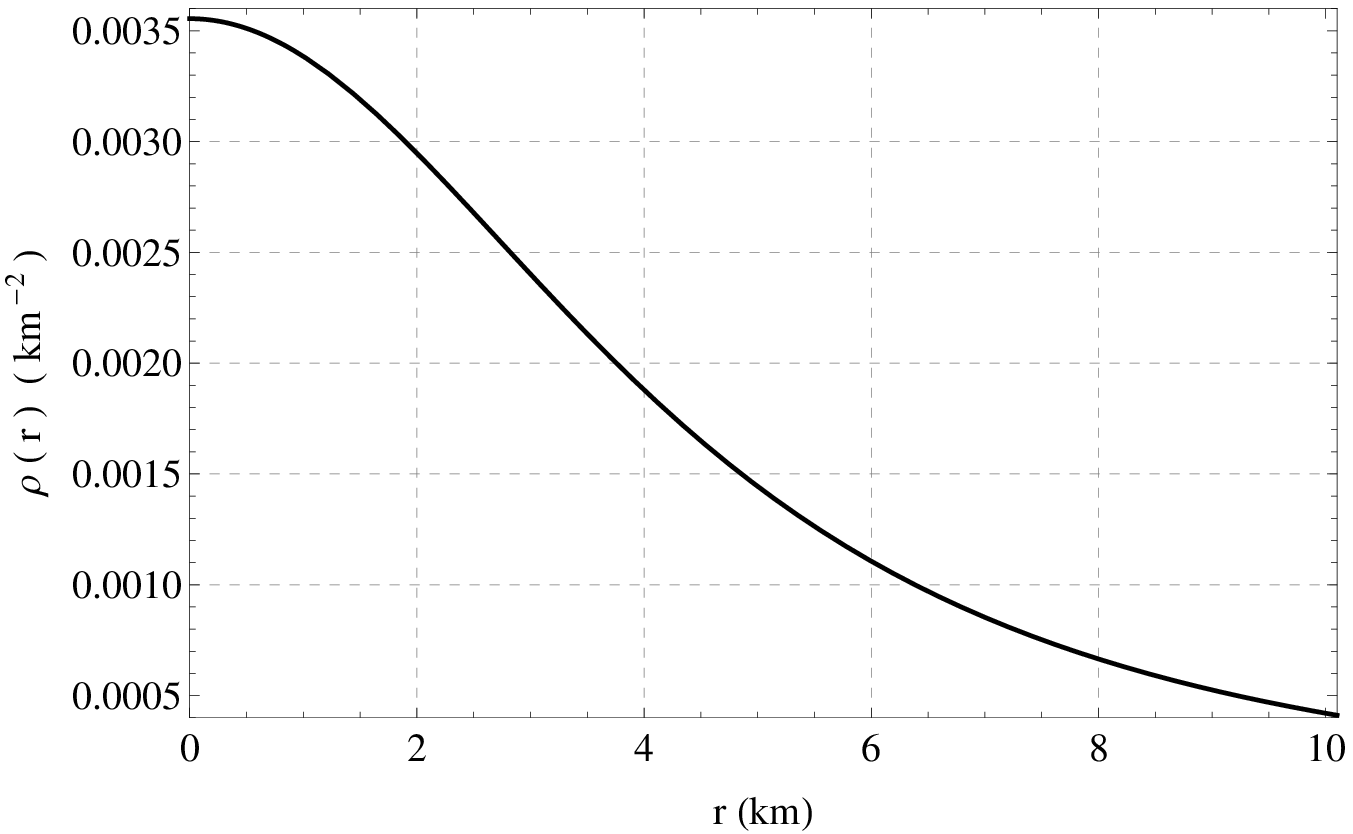}\\
\end{tabular}
\end{center}
\caption{The metric potentials are plotted against $r$ by taking $B=0.38$ $C=0.022$ and $F=10.12$ and $K=0.001$. Variation of matter density is plotted against $r$ by taking the same values of the constant mentioned earlier.}\label{md}
\end{figure*}

\begin{figure*}[thbp]
\begin{center}
\begin{tabular}{rl}
\includegraphics[width=6.8cm]{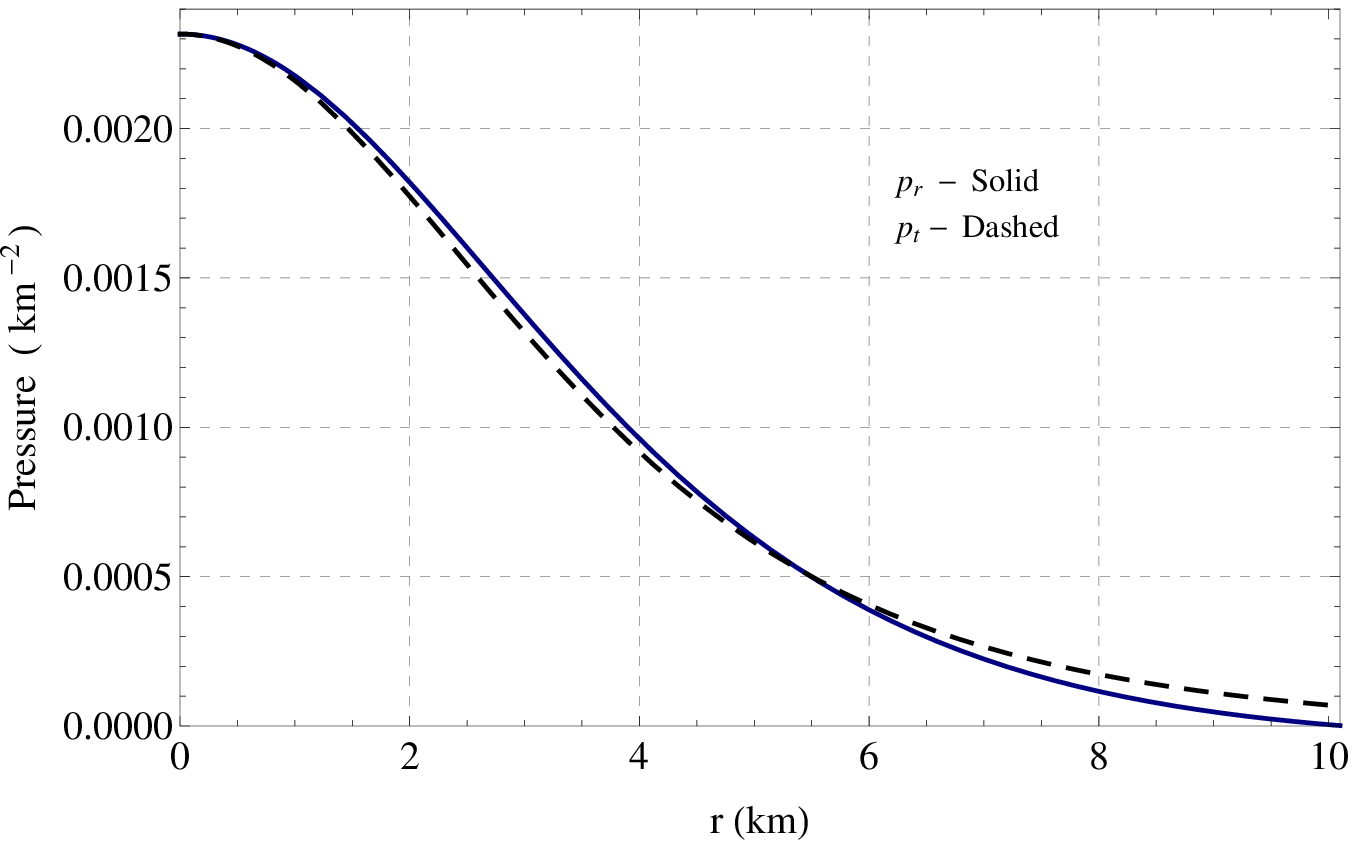}&
\includegraphics[width=6.8cm]{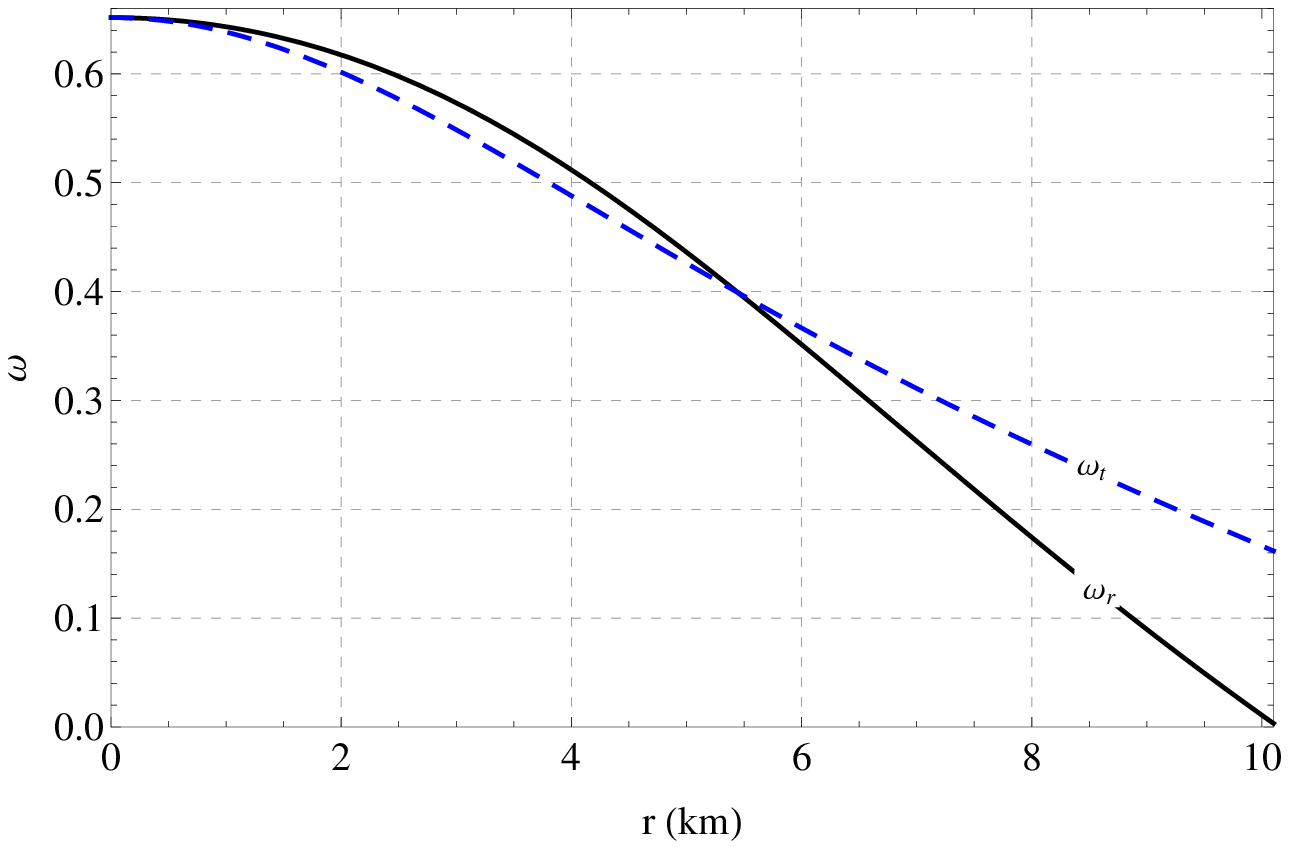}\\
\end{tabular}
\end{center}
\caption{Variation of pressures and pressure to density ratio are plotted against $r$ by taking the same values of the constant mentioned in fig. \ref{md} for charged and uncahrged configuration.}\label{prpt}
\end{figure*}

\section{Physical analysis of our present model}
Our present model satisfies the following conditions:

\begin{enumerate}
\item We know that the metric coefficients should be regular inside the stellar interior. From our solution we can easily check that $e^{\lambda(r=0)}=1$ and $e^{\nu(r=0)}=B$, a positive constant. To see the characteristic of the metric potential we plotted the graph of $e^{-\nu}$ and $e^{-\lambda}$ in fig. \ref{md}(left). The profiles show that metric coefficients are regular and monotonic decreasing function of $r$ inside the stellar interior.

\item The matter density, radial and transverse pressure should be positive inside the stellar interior for a physically acceptable model. The radial pressure should be vanish at the boundary of the star.

Moreover the central density, central pressure and Zeldovich condition can be obtained as,
\begin{eqnarray*}
\rho_c &=& \frac{6BC^{2}F}{\pi}>0\\
p_{rc} &=& p_{tc}=\frac{C(1-4BCF)}{2\pi}>0\\
{p_{rc} \over \rho_c} &=& \frac{1-4BCF}{12BCF} \le 1
\end{eqnarray*}
The above equations imply that our model is free from central singularities and it also gives a constraint on $BCF$ as $1/16 \le BCF < 1/4$.

Now for our model the gradient of matter density and radial pressure are obtained as,

\begin{eqnarray}
\frac{d\rho}{dr} &=& -\left[\frac{2 C K r}{(1 + C r^2)^2}+\frac{512 B^2 C^4 F^2 r (5 + 16 B C^2 F r^2)}{(1 + 16 B C^2 F r^2)^3}\right]\nonumber\\
\\
\frac{dp_r}{dr} &=& \frac{2 C r} {(1 + C r^2)^2 (1 + 16 B C^2 F r^2)^2}\big[K-4C \nonumber\\
&&+32BC^{2}F\{8BCF(1+2Cr^{2}+C^{2}r^{4}+CKr^{4})\nonumber\\
&&+(K-4C)r^{2}-2\} \big]
\end{eqnarray}

\begin{figure*}[thbp]
\begin{center}
\begin{tabular}{rl}
\includegraphics[width=6.8cm]{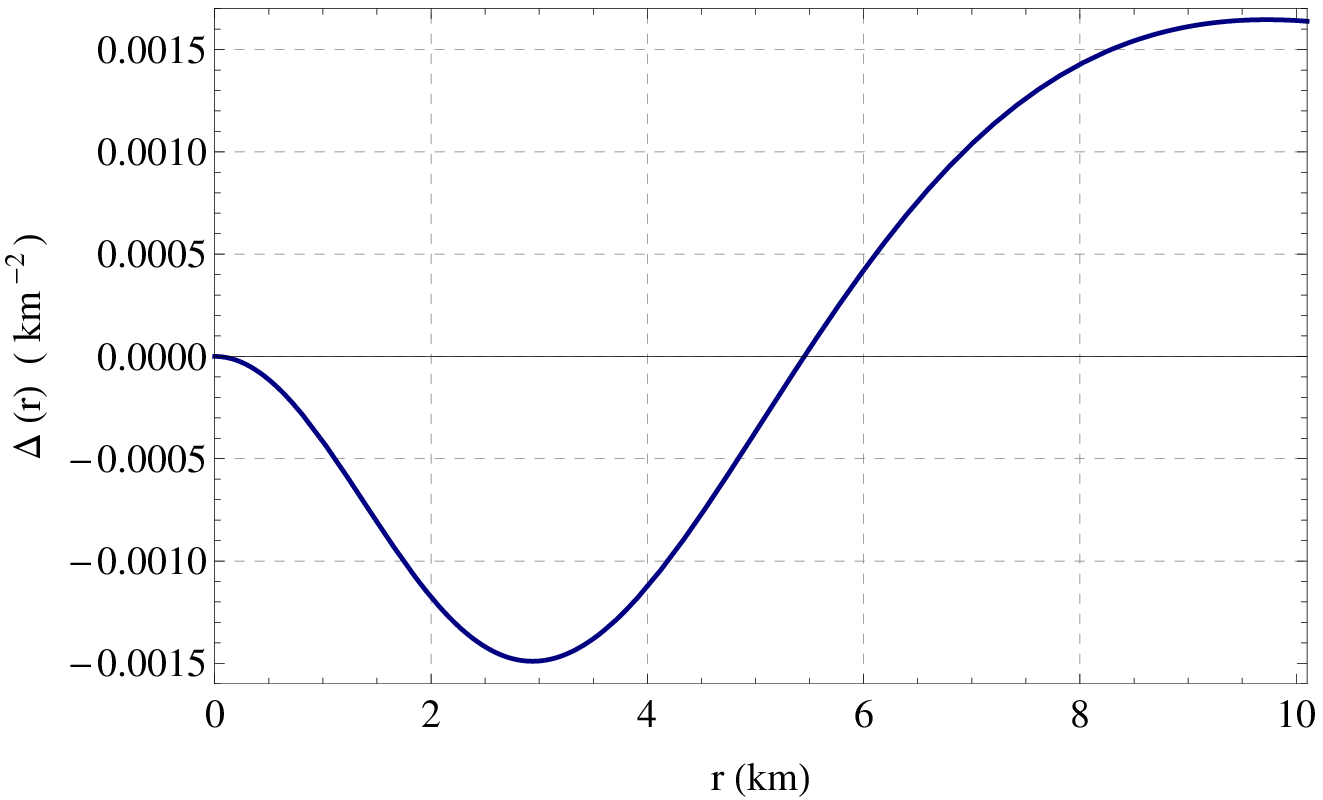}&
\includegraphics[width=6.8cm]{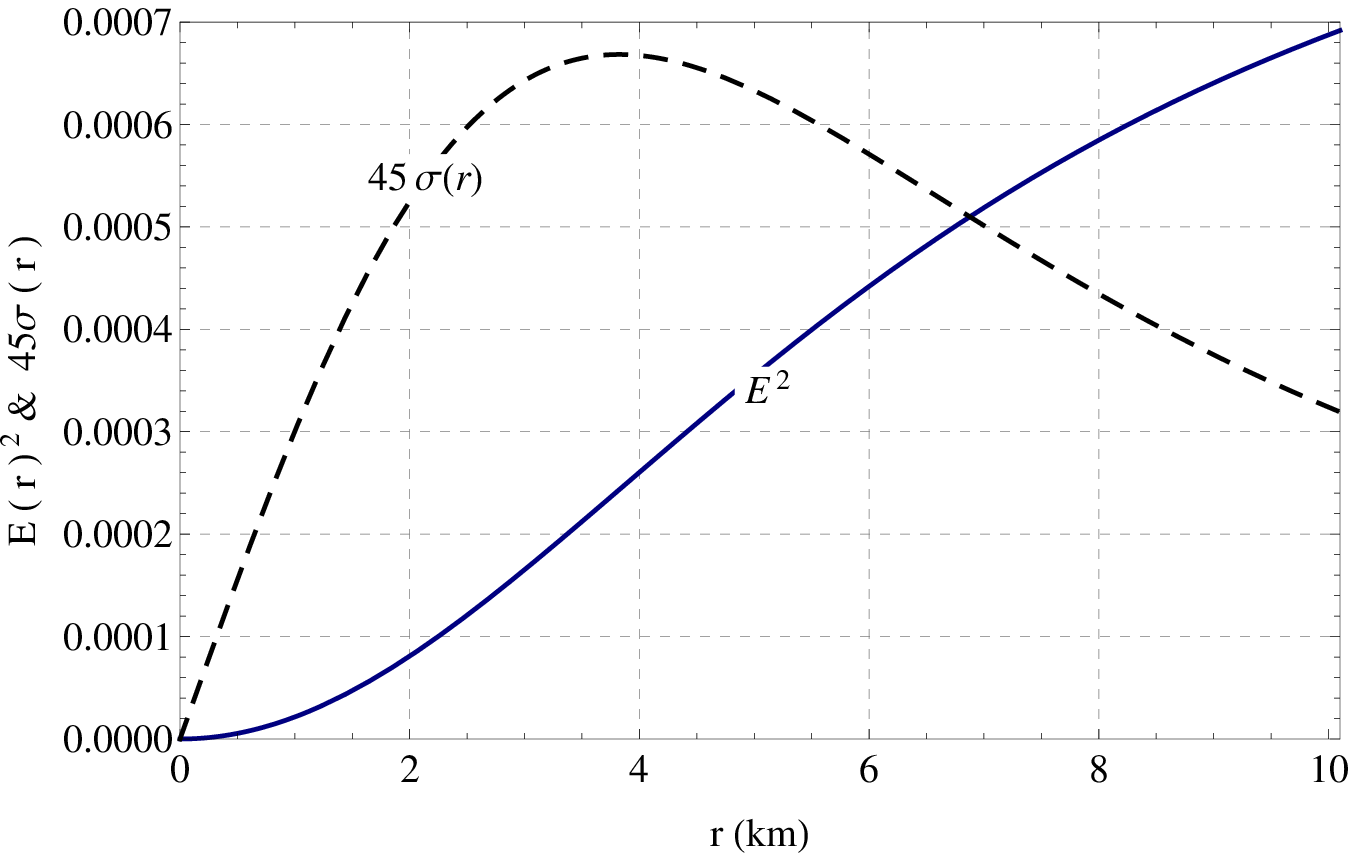}\\
\end{tabular}
\end{center}
\caption{Variation of anisotropy, $45\sigma (r)$ and electric field intensity with radial coordinates $r$ by taking the same values of the constant mentioned in fig. \ref{md}}\label{dp}
\end{figure*}

We note that at the point $r=0$ both $d\rho/dr=0$ and $dp_r/dr=0$ and,
\begin{eqnarray}
{d^{2}\rho \over dr^{2}}=-\frac{C}{4\pi}(K+1280B^{2}C^{3}F^{2})\\
\frac{d^{2}p_r}{dr^{2}}=-\frac{-CK+4C^2(1+16BCF-256B^2C^2F^2)}{4\pi}
\end{eqnarray}

The profile of matter density, radial and transverse pressure are shown in fig. \ref{md}(right) and fig. \ref{prpt} (left) respectively. The figures show that the matter density $\rho$, radial pressure $(p_r)$ and transverse pressure $(p_t)$ are monotonic decreasing function of $r$. All are positive for $0<r\leq r_b$ ($r_b$ being the boundary of the star) and both $\rho$ and $p_t$ are positive at the boundary where as the radial pressure vanishes there. The profile of the pressure to density density ratios are also monotonically decreasing and less than 1, fig. \ref{prpt} (right). The profile of $d\rho/dr$, $dp_r/dr$ and $dp_t/dr$ are plotted in fig.~5 (right). The plots show that both $d\rho/dr$, $dp_r/dr$ and $dp_t/dr$ are negative which once again verify that $\rho$, $p_r$ and $p_t$ are monotonic decreasing function of $r$.

\item The profile of the anisotropic factor $\Delta$ is shown against $r$ in fig.\ref{dp} (left). The anisotropic factor is negative (i.e. $p_t<p_r$) from center till $r=5.47 km$  and positive (i.e. $p_t>p_r$) for $r>5.47km$ till upto the surface in increasing trend., which implies $p_t>p_r$. Moreover at the center of the star the anisotropic factor vanishes which is also a required condition. The electric field vanishes at the center and monotonically increasing outward while the charge density is increasing upto $r=3.84km$ and then decreasing till the surface, fig \ref{dp} (right).\\
\end{enumerate}

\begin{figure*}[thbp]
\begin{center}
\begin{tabular}{rl}
\includegraphics[width=6.8cm]{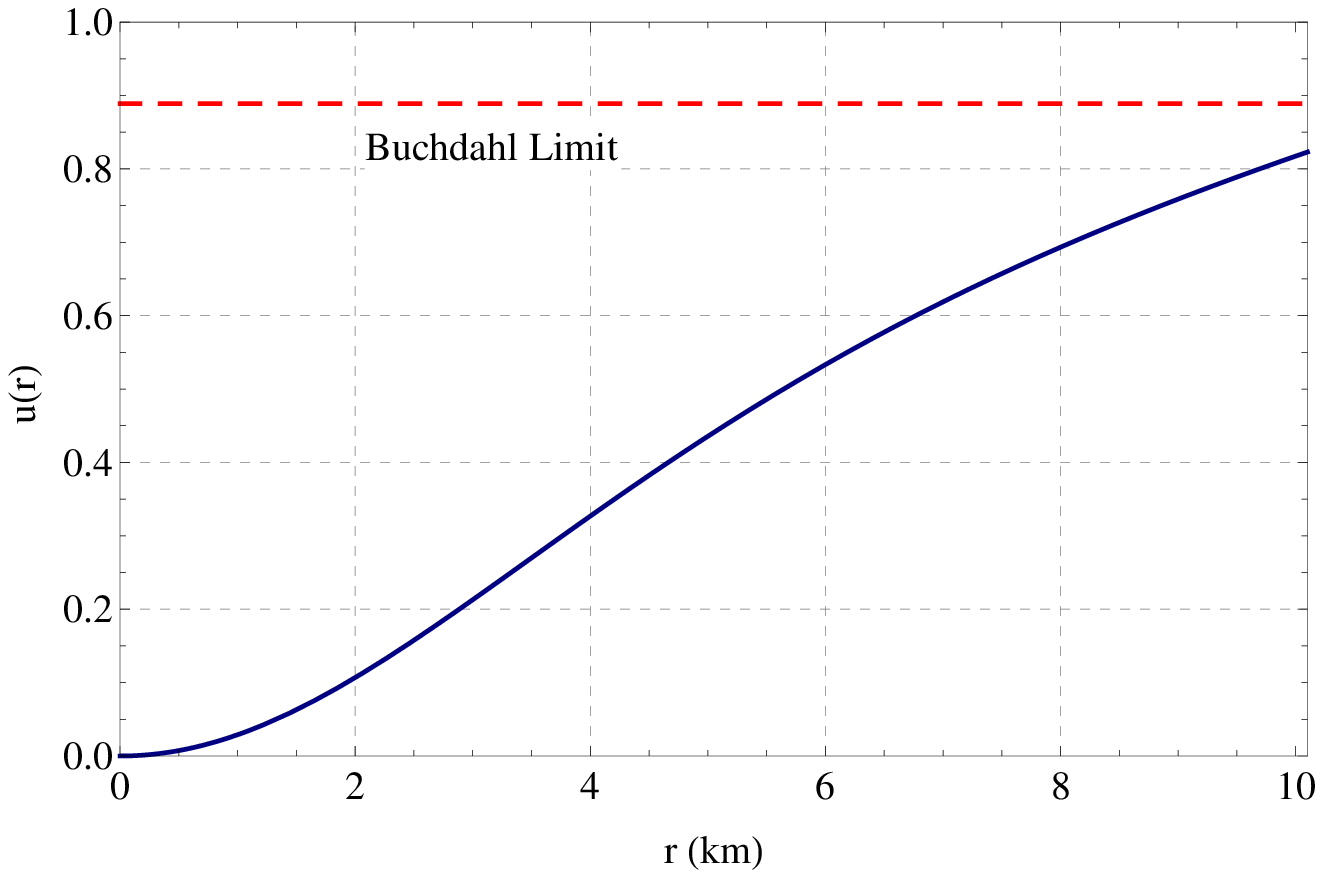}&
\includegraphics[width=6.8cm]{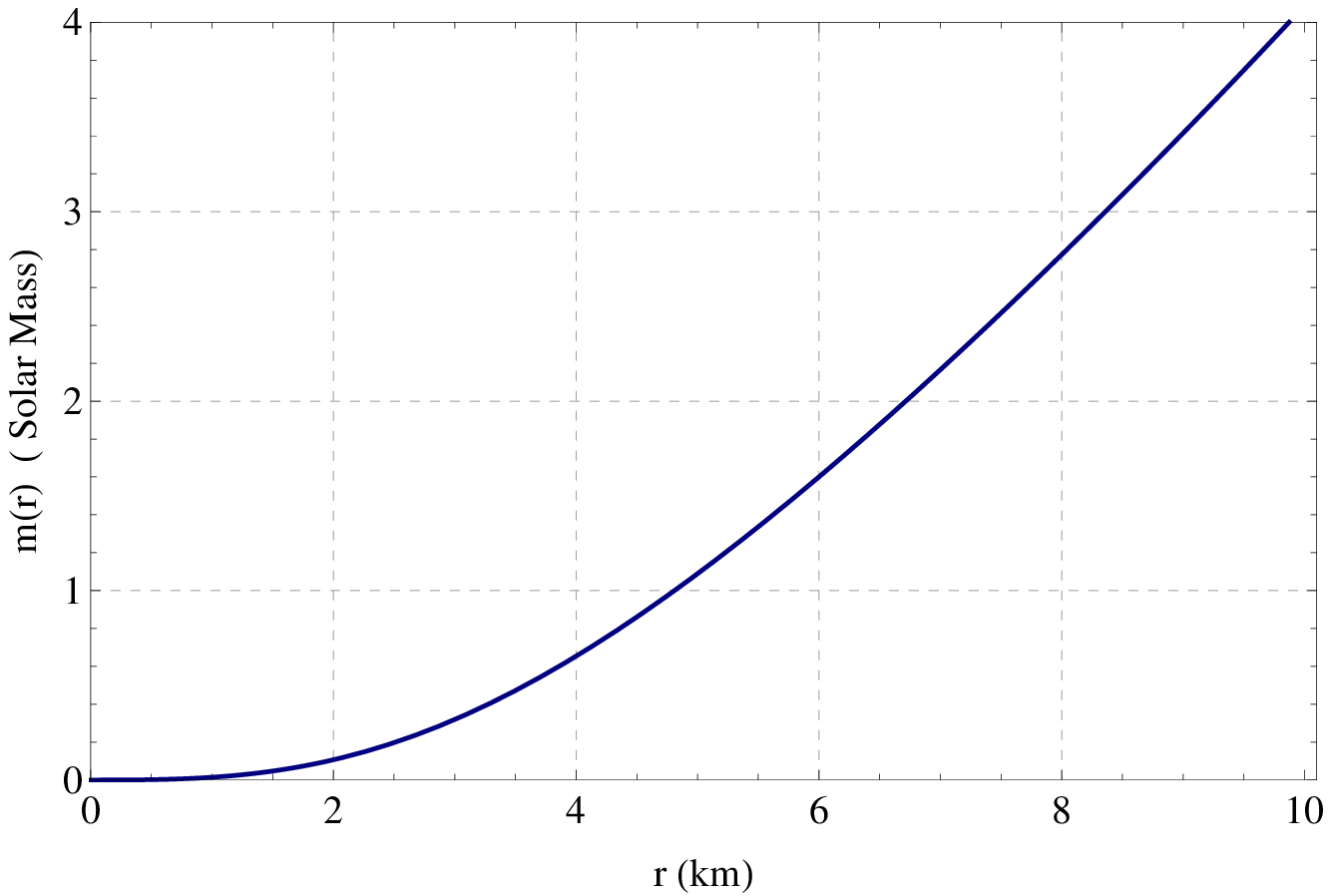}\\
\end{tabular}
\end{center}
\caption{Variation compactness parameter and mass are plotted against $r$ by taking the same values of the constant mentioned in fig. \ref{md}.}\label{delta}
\end{figure*}

\section{Mass-radius relation and compactness parameter}
The mass and compactness parameter of the compact star is obtained as,
\begin{eqnarray}
m(r) &=& \int_0^r4\pi\rho r^{2}dr\nonumber\\
&=&-\frac{K r^3}{6} + \frac{r}{2}\Big[\frac{K}{C} -\frac{16 B C^2 F r^2}{1+16 B C^2 F r^2}\Big]\nonumber\\
&& -
 \frac{K}{2C^{3/2}}tan^{-1}\sqrt{Cr}\\
u(r) &=& {2m(r) \over r}=-\frac{K r^2}{3} + \frac{K}{C} -\frac{16 B C^2 F r^2}{1+16 B C^2 F r^2}\nonumber\\
&& -\frac{K}{C^{3/2}}\frac{tan^{-1}\sqrt{Cr}}{r}
\end{eqnarray}

The profile of the compactness parameter and mass function is plotted against $r$ in fig \ref{delta}. The profile shows that mass and compact parameter are increasing function of $r$ and they are regular everywhere inside the stellar interior.

\section{Energy Conditions}

In this section we are going to verify the energy conditions namely null energy condition (NEC), dominant energy condition (DEC) and weak energy condition(WEC) at all points in the interior of a star which will be satisfied if the following inequalities hold simultaneously:
\begin{eqnarray}
\text{NEC} &:& \rho(r)\geq  0 ,\\
\text{WEC} &:& \rho(r)-p_r(r) \geq  0~~ \text{and} ~~\rho(r)-p_t(r) \geq  0,\\
\text{DEC} &:& \rho(r) \ge |p_r|,~|p_t|
\end{eqnarray}

\begin{figure*}[thbp]
\begin{center}
\begin{tabular}{rl}
\includegraphics[width=6.8cm]{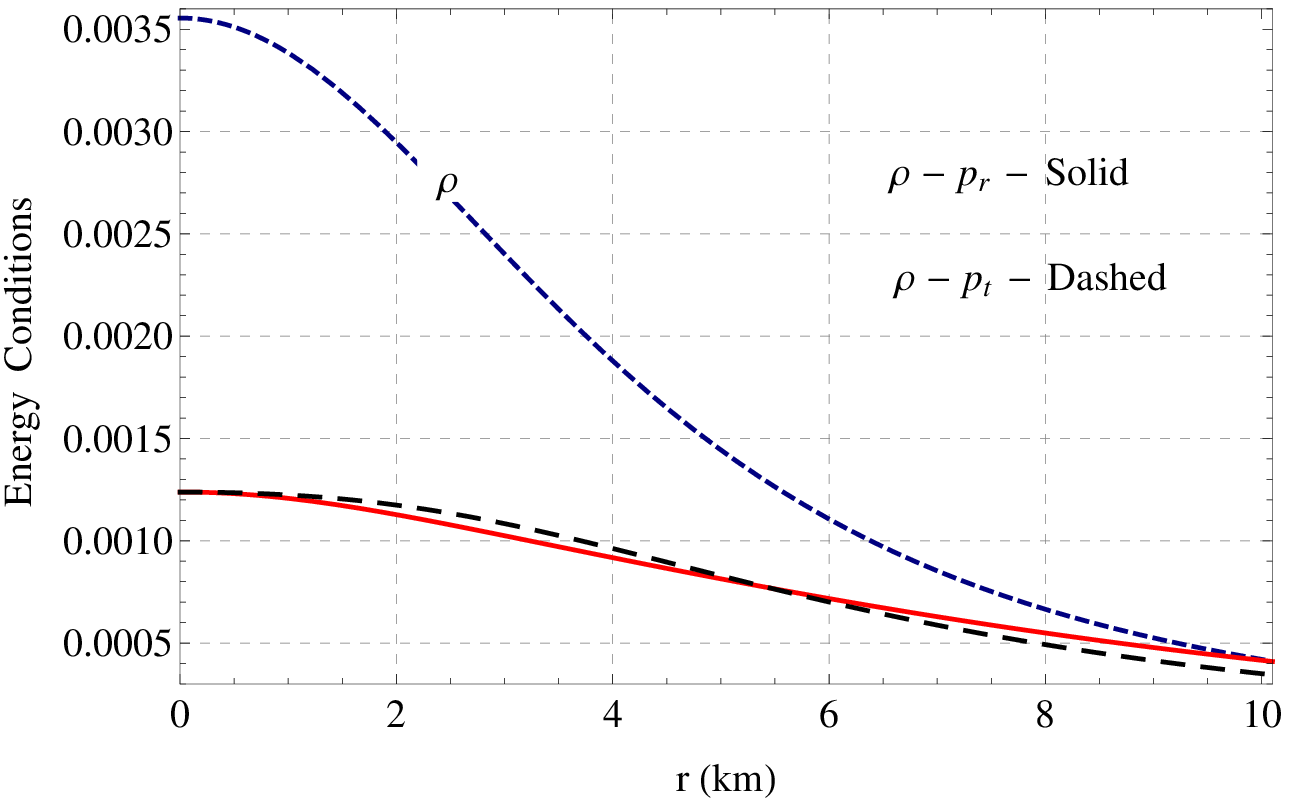}&
\includegraphics[width=6.8cm]{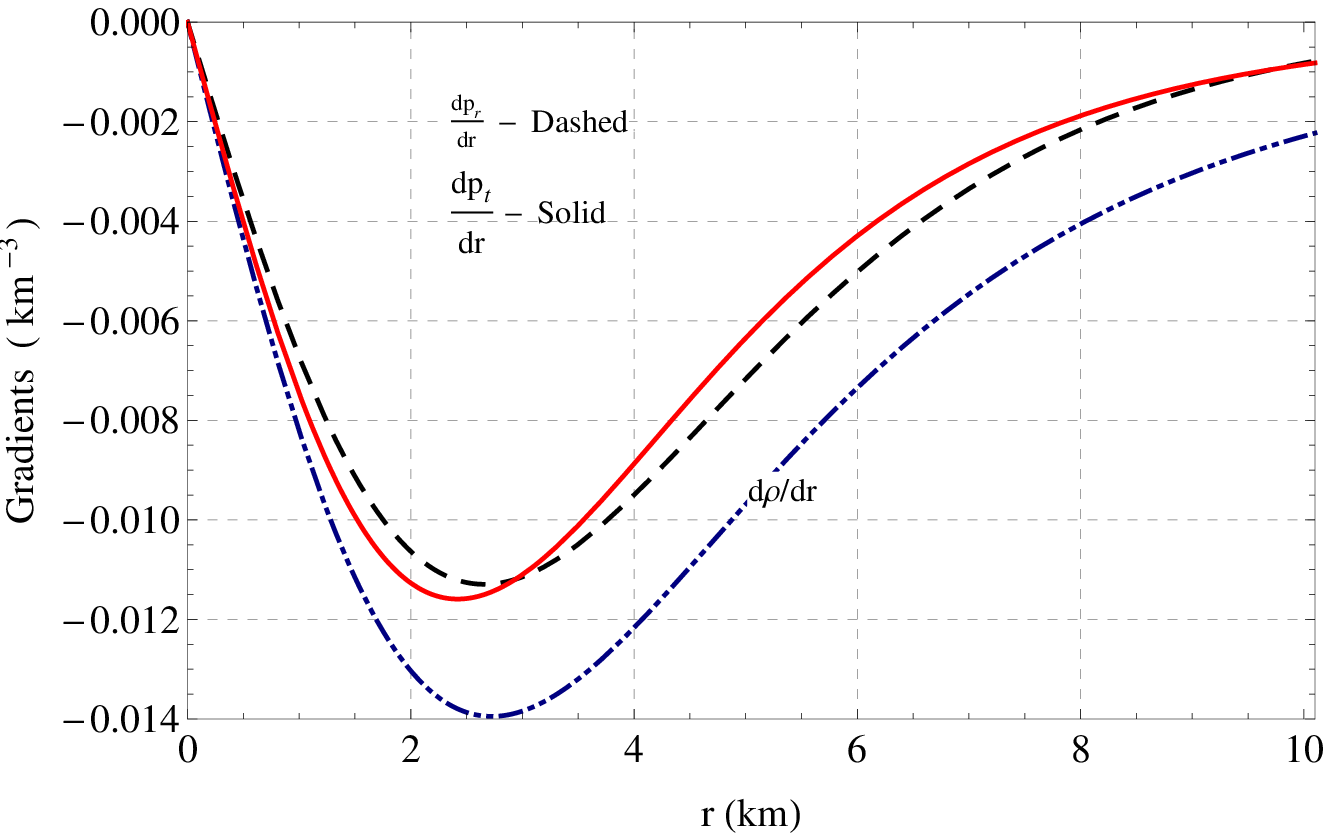}\\
\end{tabular}
\end{center}
\caption{The energy conditions, density and pressure gradients are plotted against $r$ for both charged model  and uncharged model by taking the same values of the constants mentioned in fig. \ref{md}}\label{ec}
\end{figure*}

We will check the energy conditions with the help of graphical representation. In Fig. \ref{ec} (left), we have plotted the L.H.S of the above inequalities which verifies that  all the energy conditions are satisfied at the stellar interior.

\section{Stability of the model and equilibrium}

\subsection{Equilibrium under various forces}
Equilibrium state under four forces $viz$ gravitational, hydro-statics, anisotropic and electric forces can be analyze whether they satisfy the generalized Tolman-Oppenheimer-Volkoff (TOV) equation or not and it is given by
\begin{equation}
-\frac{M_g(r)(\rho+p_r)}{r}e^{\frac{\nu-\lambda}{2}}-\frac{dp_r}{dr}+\frac{2}{r}(p_t-p_r)+\sigma(r) E(r) e^{\lambda/2}=0, \label{to1}
\end{equation}
where $M_g(r) $ represents the gravitational mass within the radius $r$, which can derived from the Tolman-Whittaker formula and the Einstein's field equations and is defined by

\begin{eqnarray}
M_g(r) &=& 4 \pi \int_0^r \big(T^t_t-T^r_r-T^\theta_\theta-T^\phi_\phi \big) r^2 e^{(\nu+\lambda)/2}dr \label{mg}
\end{eqnarray}

For the Eqs. (\ref{dens})-(\ref{prt}), the above Eq. (\ref{mg}) reduced to
\begin{equation}
M_g(r)=\frac{1}{2}re^{(\lambda-\nu)/2}~\nu'.
\end{equation}

Plugging the value of $M_g(r)$ in equation (\ref{to1}), we get
\begin{equation}
-\frac{\nu'}{2}(\rho+p_r)-\frac{dp_r}{dr}+\frac{2}{r}(p_t-p_r)+\sigma(r) E(r) e^{\lambda/2}=0.
\end{equation}

The above expression may also be written as
\begin{equation}
F_g+F_h+F_a+F_e=0,
\end{equation}
where $F_g, F_h$, $F_a$ and $F_e$ represents the gravitational, hydrostatics and anisotropic and electric forces respectively.\\

\begin{figure*}[thbp]
\begin{center}
\begin{tabular}{rl}
\includegraphics[width=6.8cm]{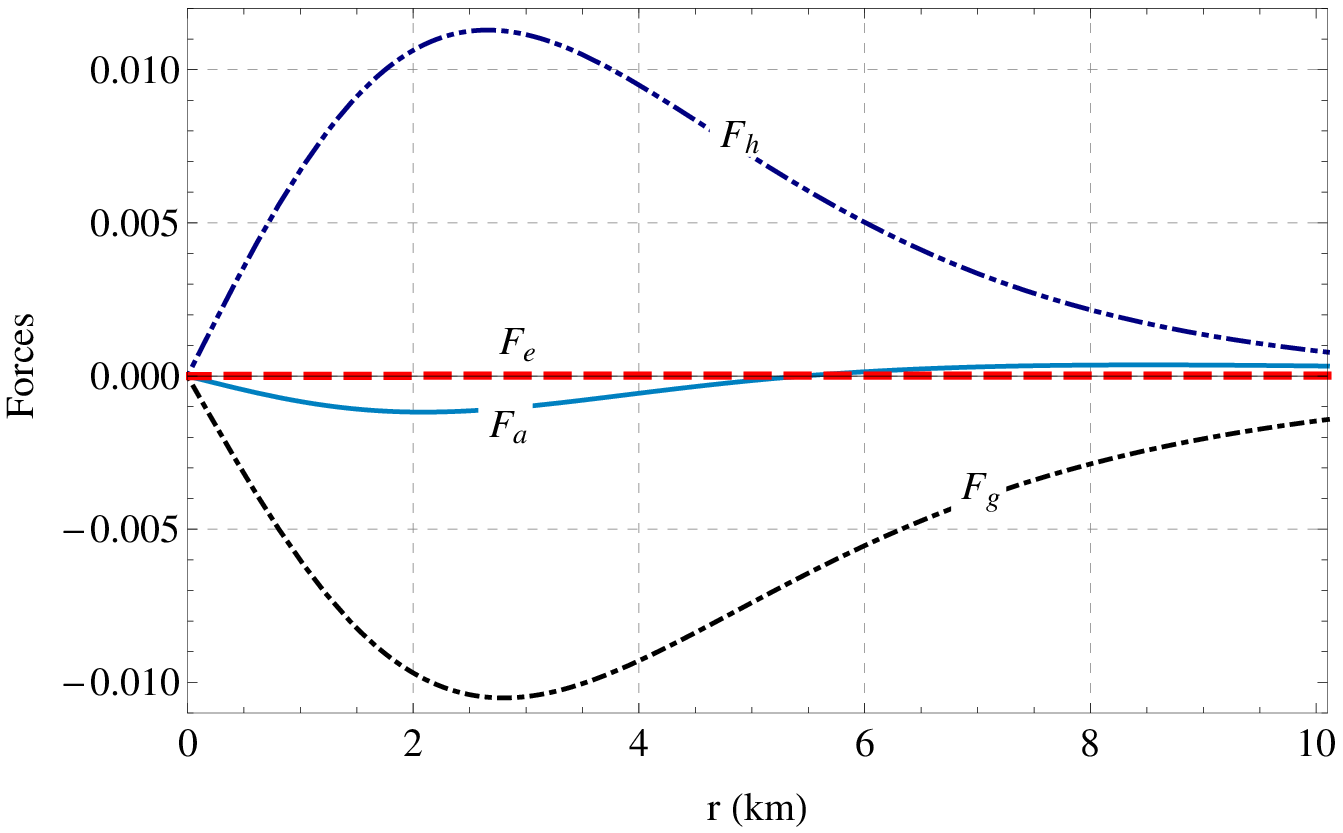}&
\includegraphics[width=6.8cm]{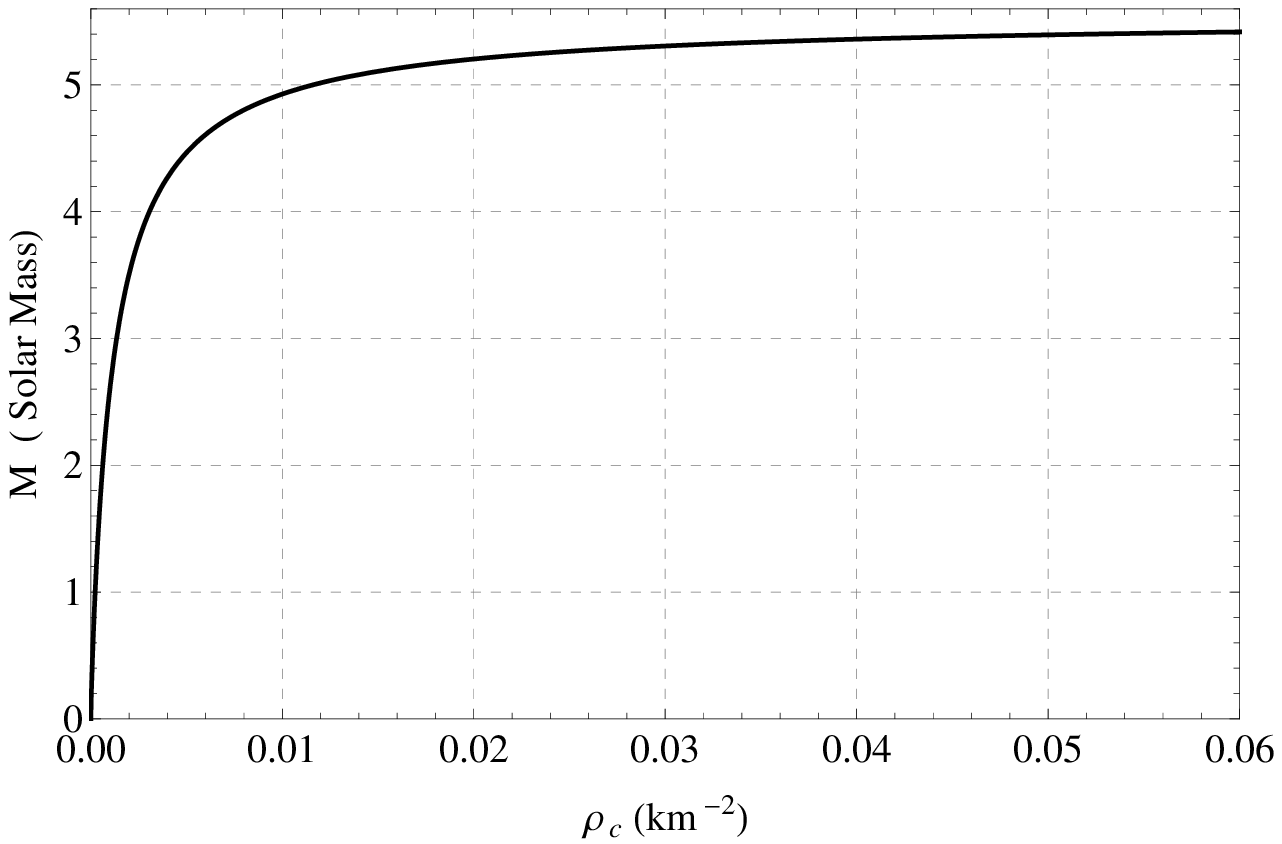}\\
\end{tabular}
\end{center}
\caption{Counter-balancing of four forces acting on the system are plotted with radial coordinates $r$ and variation of mass with central density are shown by taking the same values of the constant mentioned in fig. \ref{md}}\label{tov}
\end{figure*}

The expression for $F_g,~F_h$, $F_a$  and $F_e$ can be written as,

\begin{eqnarray}
F_g &=& -\frac{\nu'}{2}(\rho+p_r)\nonumber\\
&=&-\frac{ c^2 r \Big[1 + 8 B c F (1 + 3 c r^2)\Big]}{\pi(1 + c r^2)^2 (1 + 16 B c^2 F r^2)^2}\\
F_h &=& -\frac{dp_r}{dr}\nonumber\\
&=&-\frac{ C r\Big[K-4C+32BC^{2}F f_1(r)\Big]}{4\pi(1 + C r^2)^2 (1 + 16 B C^2 F r^2)^2}\\
F_a &=& \frac{C r^2 [-K - 16 B C^2 F f_2(r)]}{2\pi r(1 + C r^2) (1 + 16 B C^2 F r^2)^2}\\
F_e &=& \sigma Ee^{\lambda/2}=\frac{CKr(3+2Cr^{2})}{4\pi(1+Cr^{2})^{2}}
\end{eqnarray}
\begin{eqnarray*}
f_1(r)&=&8BCF(1+2Cr^{2}+C^{2}r^{4}+CKr^{4})\\
&&+(K-4C)r^{2}-2\\
f_2(r)&=&\{1 + 2 K r^2 - 8 B C F (1 + C r^2- 2 K C r^4)\}
\end{eqnarray*}
The profile of three different forces are plotted in fig. \ref{tov} (left). The figure shows that gravitational force is dominating is nature and is counterbalanced by the combine effect of hydrostatics and anisotropic force.

\subsection{Causality and stability condition}
In this section we are going to find the subliminal velocity of sound and stability condition. For a physically acceptable model of anisotropic fluid sphere the radial and transverse velocity of sound should be less than 1 which is known as causality conditions. The radial velocity $(v_{sr}^{2})$ and transverse velocity $(v_{st}^{2})$ of sound can be obtained as

\begin{eqnarray}
v_{sr}^{2} &=& \frac{(1+16 BC^2F r^2)}{K + 16 B C^2 F f_3(r)}\times\bigg[4C-K-32BC^{2}F\times \nonumber\\&&\Big\{8BCF(1+2Cr^{2}+C^{2}r^{4}+CKr^{4})\nonumber\\
&&+(K-4C)r^{2}-2\Big\}\bigg] \\
v_{st}^{2} &=& \frac{1}{K + 16 B C^2 F f_3(r)}\times \bigg[K + 4C\Big[1 + 4 B C F \times \nonumber\\
&&\Big\{6 + 3 (4 C + K) r^2 + 256 B^2 C^4 F^2 K r^6 \nonumber\\
&&+ 16 B C F (2C^2r^4 + 3C K r^4- 2 C r^2-2)\Big\}\Big]\bigg]
\end{eqnarray}

\begin{eqnarray*}
f_3(r)&=&[3 K r^2 +256 B^2 C^3 F^2 r^2 \{1 + 2 C r^2 + C(C + K) r^4\} \\
&&+16 B C F \{5 + 10 C r^2 + C (5 C + 3 K) r^4\}]
\end{eqnarray*}
\begin{figure*}[thbp]
\begin{center}
\begin{tabular}{rl}
\includegraphics[width=6.8cm]{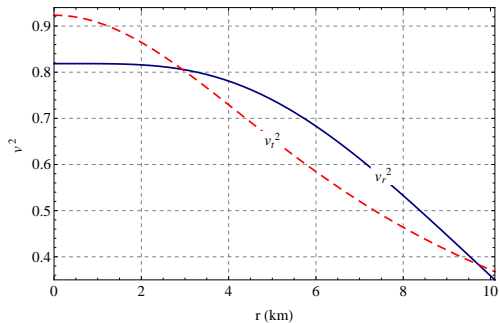}
\end{tabular}
\end{center}
\caption{Square of sound velocities is plotted against $r$ by employing the same values of the arbitrary constants as mentioned in fig. \ref{md}.}\label{sv}
\end{figure*}

The profile of radial and transverse velocity of sound have been plotted in fig. \ref{sv}, the figure indicates that our model satisfies the causality condition.

\subsection{Adiabatic index and stability condition}

For a relativistic anisotropic sphere the stability is related to the adiabatic index $\Gamma$, the ratio of two specific heats, defined by \cite{chan},
\begin{equation}
\Gamma=\frac{\rho+p_r}{p_r}\frac{dp_r}{d\rho}; ~~  \Gamma_t = \frac{\rho+p_t}{p_t}\frac{dp_t}{d\rho}
\end{equation}

Now $\Gamma>4/3$ gives the condition for the stability of a Newtonian sphere and $\Gamma =4/3$ being the condition for a neutral equilibrium proposed by \cite{bondi64}. This condition changes for a relativistic isotropic sphere due to the regenerative effect of pressure, which renders the sphere more unstable. For an anisotropic general relativistic sphere the situation becomes more complicated, because the stability will depend on the type of anisotropy. For an anisotropic relativistic sphere the stability condition is given by \cite{chan},

\begin{equation}
\Gamma>\frac{4}{3}+\left[\frac{4}{3}\frac{(p_{ti}-p_{ri})}{|p_{ri}^\prime|r}+\frac{1}{2}\kappa\frac{\rho_ip_{ri}}{|p_{ri}^\prime|}r\right],
\end{equation}
where, $p_{ri}$, $p_{ti}$, and $\rho_i$ are the initial radial, tangential, and energy density in static equilibrium satisfying (\ref{to1}). The first and last term inside the square brackets represent the anisotropic and relativistic corrections respectively and both the quantities are positive that increase the unstable range of $\Gamma$ \citep{her1,chan}.

\begin{figure*}[thbp]
\begin{center}
\begin{tabular}{rl}
\includegraphics[width=6.8cm]{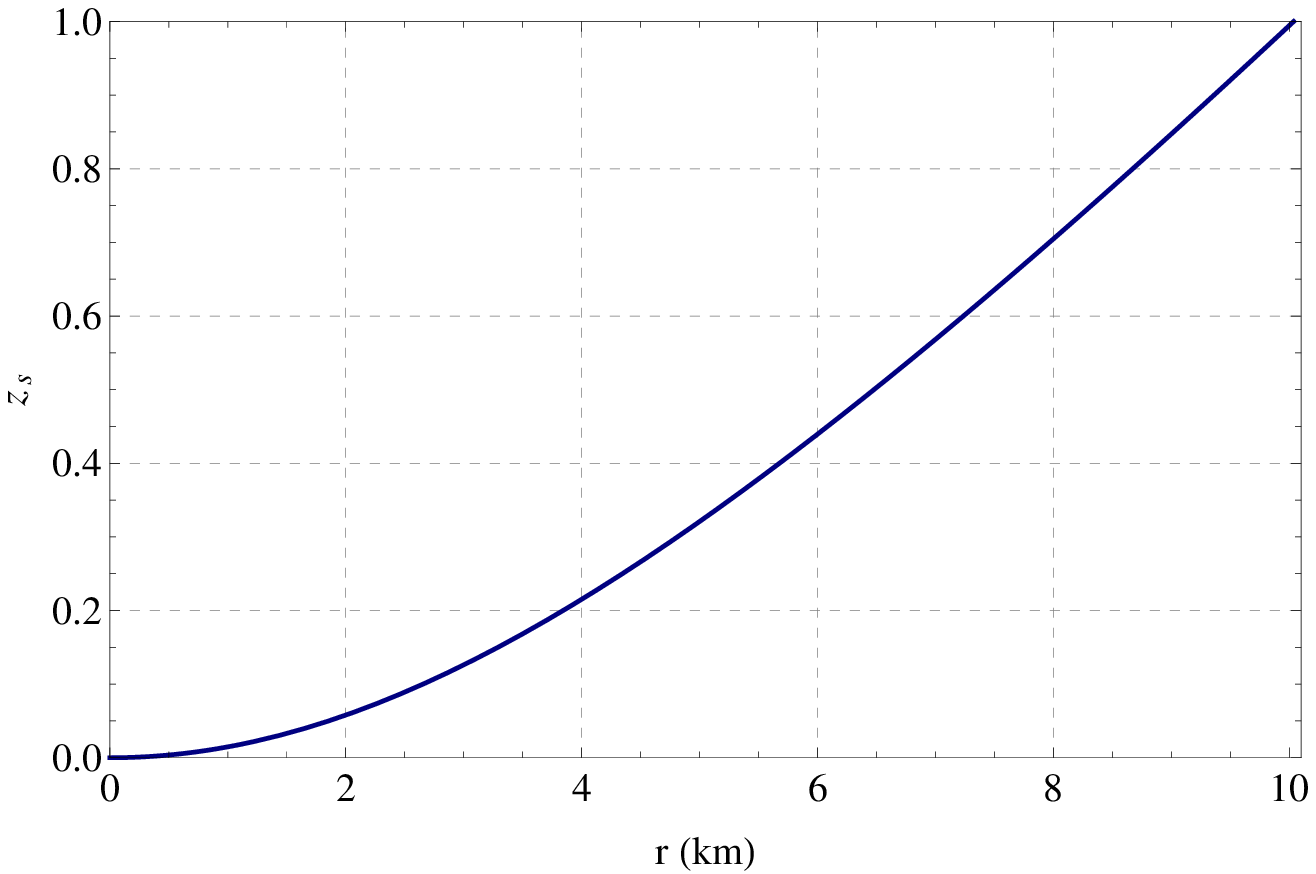}&
\includegraphics[width=6.8cm]{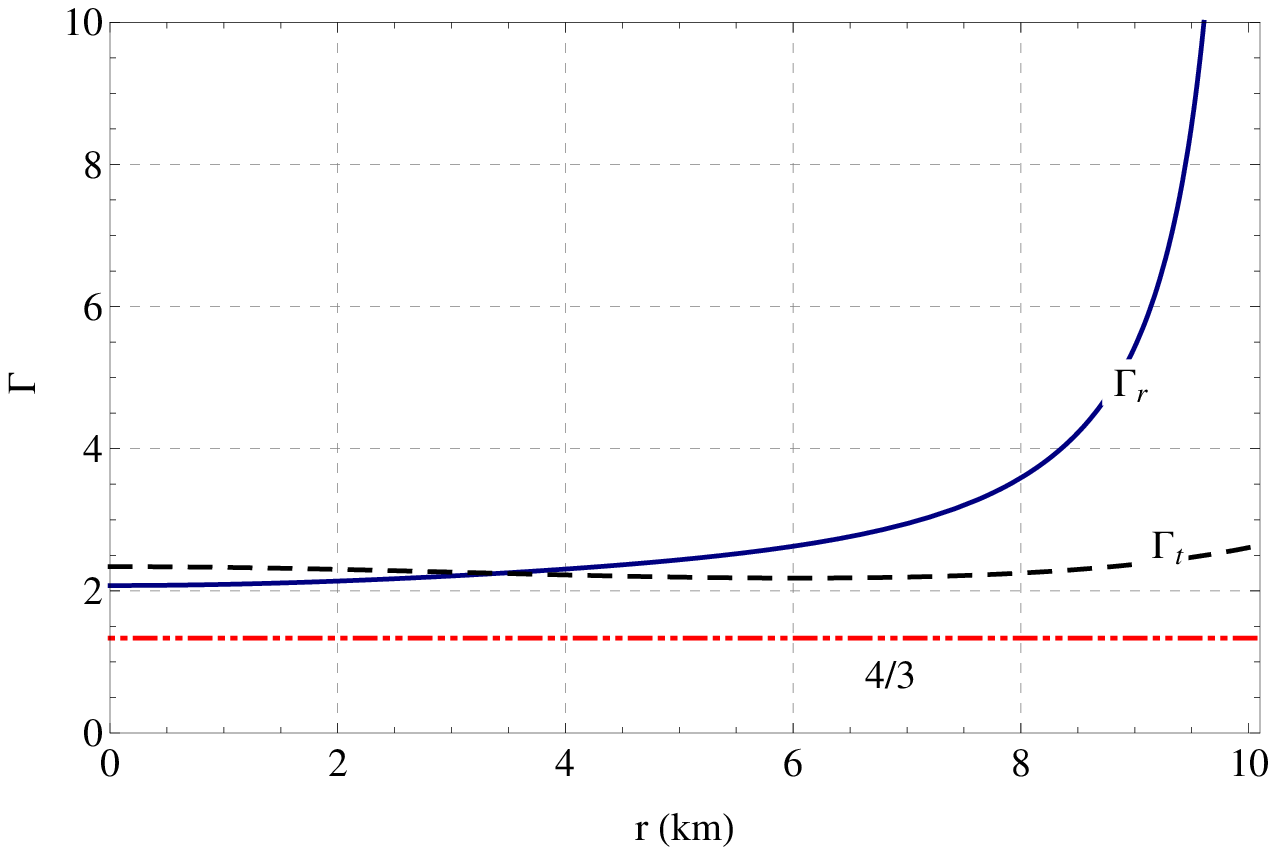}\\
\end{tabular}
\end{center}
\caption{Surface red-shift and relativistic adiabatic index are plotted against $r$ by employing the same values of the arbitrary constants as mentioned in fig. \ref{md}.}\label{re}
\end{figure*}

\subsection{Harrison-Zeldovich-Novikov static stability criterion}

The stability analysis adopted by \cite{chan64}, \cite{har65} etc. requires the determination of eigen-frequencies of all the fundamental modes. However, \cite{har65} and \cite{zel} simplify such messy calculations and reduced it to much more simpler formulism. They have assumed that the adiabatic index of a pulsating star is same as in a slowly deformed matter. This leads to a stable configuration only if the mass of the star is increasing with central density i.e. $dM/d\rho_c > 0$ and unstable if $dM/d\rho_c < 0$.

In our solution, the mass as a function of central density can be written as
\begin{eqnarray*}
M &=& \frac{8\pi \rho_c R^{3}/6}{1+ 16\pi \rho_c R^2/3}+{K \alpha R^5 \over 2(1+\alpha r^2)}\\
\alpha&=&\sqrt{{\pi \rho_c \over 6BF}} \label{mrho1}
\end{eqnarray*}
which gives us (for a given radius, $B$ and $F$)

\begin{eqnarray}
{d M \over d \rho_c} &=& \frac{3 R^3 \sqrt{\pi  \rho_c /B F} }{2 \rho_c \big(8 \pi  \rho_c R^2+3\big)^2 \big[R^2 \sqrt{6 \pi  \rho_c/B F}+6\big]^2}\times \nonumber\\
&&\Big[R^2 \Big\{48 \pi  \rho_c \Big(R^2 \sqrt{\pi \rho_c/B F}+2 \sqrt{6}\Big)\nonumber\\
&&+\sqrt{6} K \Big(8 \pi  \rho_c R^2+3\Big)^2\Big\}+288 \sqrt{\pi  \rho_c B F}\Big] \label{dm}
\end{eqnarray}
Since all the quantities used in (\ref{dm}) are positive finite values, $dM/d\rho_c$ is always $>0$, which implies that the total mass of stellar system increases with increase in central density. This condition can be further confirm by Fig. \ref{tov} (right).

\section{Discussion and conclusion}

It has been observed that the physical parameters $\big(e^{-\lambda},~e^{-\nu},~\rho,~p_r,~p_t,~p_r/\rho,~p_t/\rho,~v_r^2,~v_t^2 \big)$ are positive at the center, within the limit of realistic equation of state and monotonically decreasing outward (Figs. \ref{md}, \ref{prpt}, \ref{sv}). However, the anisotropy, $z_s,~E^2$ and $\Gamma$ are increasing outward which is necessary for a physically viable configuration (Figs. \ref{dp}, \ref{re}). The proper charged density at the interior is also shown in Fig. \ref{dp}. The variation of compactness parameter $u$ and the mass distribution with radial coordinates in Figs. \ref{delta} signifies that the new charged anisotropic  solution leads to very stiff EoS. This stiff EoS yields a compactness parameter of 0.823, which is very closed to the Buchdahl limit 0.889 and also the total mass of  $4.156M_\odot$ is bounded in a very small radius of 10.1$km$ only. For the chosen values mentioned in Fig. \ref{md}, the variation of anisotropy factor signifies that for $0\le r \le 5.45km$, $\Delta < 0$ (or $p_r>p_t$) and for $5.45km<r\le 10.1km$,  $\Delta > 0$ (or $p_t>p_r$). The non-singular properties of the solution can be represent by the finite values of central density $4.7\times 10^{15}g/cm^3$, central pressure $28.03\times 10^{35}~dyne/cm^2$, central sound speed $v_r^2=0.819$ and $v_t^2=0.923$. The relativistic adiabatic index at the center are $\Gamma_{r0}=2.062$ and $\Gamma_t=2.348$ and are also increasing monotonically outward. The redshift of the configuration at the surface is of about 1.01.\\

Furthermore, our presented solution satisfies Weak Energy Condition (WEC), Null Energy Condition (NEC) and Dominant Energy Condition (DEC) which is shown in Fig. \ref{ec}. The stability factor $|v_t^2-v_r^2|$ lies in between 0 and 1 which represents a stable configuratio (Fig. \ref{sv}). The decreasing nature of pressures and density is further justified by their negativity of their gradients, Fig. \ref{ec}. The solution also represents a static and equilibrium configuration as the force acting on the fluid sphere is counter-balancing each other. For a charged anisotropic stellar fluid in equilibrium the gravitational force, the hydro-static pressure, the Coulomb force and the anisotropic force are acting through a generalized TOV-equation and they are counter-balancing to each other, Fig. \ref{tov}. The stability analysis of the solution is also extended by adopting the Harrison-Zeldovich-Novikov static stability criterion. According to the static stability criterion, the variation of mass must be increasing in trend with the increase in central density i.e. $dM/d\rho_c>0$ for stable and $dM/d\rho_c \le 0$ for unstable configurations. The Fig. \ref{tov}, we have plotted the mass by varying $\rho_c$ from $0-8.082 \times 10^{16}~g/cm^3$ and it is noticeable that the maximum mass becomes saturated at $5.418M_\odot$  from about $6.735 \times 10^{16}~g/cm^3$. Hence from all the above analysis, we conclude that the presented solution satisfy (i) TOV-equation showing its equilibrium  condition and also satisfies the static stable criterion of Harrison-Zeldovich-Novikov i.e. $dM/d\rho_c>0$.

\subsection*{Acknowledgments}
 FR would like to thank the authorities of the Inter-University Centre
for Astronomy and Astrophysics, Pune, India for providing research facilities. FR is also
thankful to DST and SERB,  Govt. of India for providing financial support.

\end{document}